\documentclass[prd,tightenlines,nofootinbib,superscriptaddress,11pt]{revtex4}

\usepackage{amsfonts,amssymb,amsthm,bbm,amsmath}
\usepackage{color,psfrag}
\usepackage[dvips]{graphicx}

\newcommand{\cC}{{\mathcal C}}

\newcommand{\cF}{{\mathcal F}}
\newcommand{\cG}{{\mathcal G}}

\newcommand{\cJ}{{\mathcal J}}
\newcommand{\cK}{{\mathcal K}}

\newcommand{\cM}{{\mathcal M}}
\newcommand{\cN}{{\mathcal N}}

\newcommand\U{{\mathrm U}}
\newcommand\C{{\mathbb C}}

\newcommand\R{{\mathbb R}}

\newcommand{\braaket}[2]{\langle #1 | #2 \rangle }

\newcommand{\braaaket}[3]{\langle #1 | #2 | #3 \rangle }
\newcommand{\bra}[1]{\langle #1  |}
\newcommand{\ket}[1]{| #1 \rangle}
\newcommand{\ketr}[1]{| #1 ]}
\newcommand{\brar}[1]{[ #1 |}

\newcommand\bz{\bar{z}}

\newcommand{\braaketrt}[2]{[ #1  |   #2 \rangle }
\newcommand{\braakettr}[2]{\langle #1  |   #2 ] }
\newcommand\one{\mathbb{I}}

\newcommand{\SU}{\mathrm{SU}}
\newcommand{\SL}{\mathrm{SL}}
\newcommand{\GL}{\mathrm{GL}}

\newcommand{\su}{{\mathfrak{su}}}

\renewcommand{\sl}{{\mathfrak{sl}}}

\newcommand{\be}{\begin{equation}}
\newcommand{\ee}{\end{equation}}
\newcommand{\beq}{\begin{eqnarray}}
\newcommand{\eeq}{\end{eqnarray}}
\newcommand{\bes}{\begin{eqnarray}}
\newcommand{\ees}{\end{eqnarray}}
\newcommand{\bea}{\begin{eqnarray}}
\newcommand{\eea}{\end{eqnarray}}
\newcommand{\nn}{\nonumber}

\newcommand{\mat} [2] {\left ( \begin{array}{#1}#2\end{array} \right ) }

\newcommand{\la}{\langle}
\newcommand{\ra}{\rangle}

\newcommand{\tr}{{\mathrm Tr}}
\newcommand{\f}{\frac}
\newcommand{\tl}{\widetilde}
\def\arr{\rightarrow}
\def\pp{\partial}

\newcommand{\eps}{\epsilon}

\newcommand{\bzeta}{\bar{\zeta}}

\newcommand{\vJ}{\vec{J}}
\newcommand{\vK}{\vec{K}}

\newcommand{\vs}{\vec{\s}}

\newcommand{\vsigma}{\vec{\sigma}}
\newcommand{\id}{\mathbb{I}}

\def\tu{\tilde{u}}
\def\tt{\tilde{t}}
\def\tz{\tilde{z}}

\def\tphi{\tilde{\phi}}

\newtheorem{theoremp}{Theorem}[]
\newtheorem{prop}[theoremp]{Proposition}

\newcommand{\fourmat}[4]{\begin{pmatrix} #1 & #2 \\ #3 & #4  \end{pmatrix}}
\newcommand{\twovec}[2]{\begin{pmatrix} #1 \\ #2 \end{pmatrix}}
\newcommand{\dualtwovec}[2]{\begin{pmatrix} #1, & #2 \end{pmatrix}}

\newcommand{\Group}[4]{\frac{ \ket{#1}\brar{#2} - \ketr{#3}\bra{#4}  }{\sqrt{\braaket{#3}{#1}\braaket{#4}{#2}  }}}

\newcommand{\s}{\sigma}

\def\tphi{\tilde{\phi}}

\newcommand{\T}{{\mathbbm T}}
\newcommand{\slc}{\SL(2,\C)}
\newcommand{\g}{\gamma}
\newcommand{\Ref}[1]{(\ref{#1})}

\begin{document}

\title{Twistor Networks and Covariant Twisted Geometries}

\author{{\bf Etera R. Livine}}
\affiliation{Laboratoire de Physique, ENS Lyon, CNRS-UMR 5672, 46 All\'ee d'Italie, Lyon 69007, France}
\affiliation{Perimeter Institute, 31 Caroline St N, Waterloo ON, Canada N2L 2Y5}
\author{{\bf Simone Speziale}}
\affiliation{Centre de Physique Th\'eorique\footnote{Unit\'e Mixte de Recherche (UMR 6207) du CNRS et des Universites Aix-Marseille I, Aix-Marseille II et du Sud Toulon-Var. Laboratoire affili\'e \`a la FRUMAM (FR 2291).}, CNRS-Luminy Case 907, 13288 Marseille Cedex 09, France}
\author{{\bf Johannes Tambornino}}
\affiliation{Laboratoire de Physique, ENS Lyon, CNRS-UMR 5672, 46 All\'ee d'Italie, Lyon 69007, France}

\date{\today}

\begin{abstract}

We study the symplectic reduction of the phase space of two twistors to the cotangent bundle of the Lorentz group. 
We provide expressions for the Lorentz generators and group elements in terms of the spinors defining the twistors.
We use this to define twistor networks as a graph carrying the phase space of two twistors on each edge.
We also introduce simple twistor networks, which provide a classical version of the simple projected spin networks living on the boundary Hilbert space of EPRL/FK spin foam models.
Finally, we give an expression for the Haar measure in terms of spinors.

\end{abstract}

\maketitle

\section{Introduction}

Although loop quantum gravity is a continuum theory, truncating it to a single graph provides a useful approximation
for regimes adapted to the coarseness of the graph \cite{twisted3,lectures}.
The truncation corresponds to capturing only a finite number of degrees of freedom of the gravitational field.
These finite degrees of freedom have been shown to correspond to the quantization of twisted geometries \cite{twisted1},
that is a notion of discrete geometries given by a collection of polyhedra associated to the cellular decomposition dual to the graph, with non-trivial extrinsic curvature among them. 
This picture can be elegantly described purely in terms of spinors and twistors \cite{twisted2}.
The mechanism is a symplectic reduction from a space of two spinors, or a single twistor, on each edge of the graph,
by a constraint that geometrically imposes the matching of the area on the face shared by the adjacent polyhedra.
The spinorial description\footnote{The relevant phase space for SU(2) spin networks is a collection of cotangent bundles of SU(2), that is the symplectic manifold $T^*\SU(2)$, with its canonical Poisson algebra, attached to each edge of the graph. Upon quantization, this turns into the Hilbert space $L_2[\SU(2),{\rm d}\mu_{\rm Haar}]$ with the angular momentum operator algebra.
Analogously, the relevant classical space for covariant spin networks is $T^*\slc$ with its canonical Poisson algebra.
In \cite{twisted2}, it was shown that $T^*\SU(2)$ can be obtained starting from the twistor space $\T\cong \C^4$, imposing the constraint that the twistor has null helicity. This leads to a description of spin networks in terms of twistors that was studied in \cite{spinor2}. The result is a new picture of a spin network as a collection of spinors on each half-edge, subject to a U(1) area-matching constraint on each edge, and SU(2)-gauge-invariance on each vertex.
} leads to a notion of spinor networks \cite{spinor2} and to a number of applications \cite{un1,un2,spinor1}.

For dynamical purposes, it is also useful to consider spin networks for the full Lorentz group \cite{freidel_livine_spinnets,projected,lift}. 
Then a generalized decomposition takes place, with pairs of twistors instead of pairs of spinors, and \emph{covariant} twisted geometries, with bivectors associated to the faces instead of vectors.
In this paper we study some details of this generalization.
First, we show that the (16-dimensional) phase space of two twistors reduces to the (12-dimensional) cotangent bundle of $\slc$, under imposition of a complex constraint among the scalar products of the twistors.
We provide an explicit parameterization of the algebra and group elements of $T^*\slc$ in terms of the spinors defining the twistors. This result generalizes the SU(2) construction of \cite{twisted2}, and coincides with what reported in \cite{Wolfgang}.

This allows us to define \emph{twistor networks}, a generalization of the spinor networks studied in \cite{spinor1}.
In particular, we identify the special class of twistor networks that corresponds to the Dupuis-Livine class of simple projected spin networks \cite{lift} which satisfy the simplicity constraints and live in the boundary of the EPRL/FK spin foam models \cite{lift,covariance}.

Our results provide new tools for the study of covariant properties of loop quantum gravity and spin foams. As first applications, in this paper we give an explicit parameterization of the holonomy in terms of the bivectors living at start and end points, similar to the one used in twisted geometries, and provide a formula for the Haar measure on $\slc$ in terms of spinors.

\section{Twistor Phase Space Representation of $T^*\slc$}

Our starting point is the phase space of a twistor, $\T\cong \C^4$, with its canonical Poisson brackets.
That is, fixing once and for all an origin, we identify a twistor $\psi$ with a pair of spinors,\footnote{Notice that although in this paper we deal with classical physics, we use Dirac's notation for vectors. This is convenient to avoid clogging the formulas with indices.
A spinor $\ket{t}$ has components $t^A$, $A=0,1$, a conjugate $\bra{t} := \dualtwovec{\bar{t^0}}{\bar{t^1}}$, and a dual
\beq\label{dual}
\ketr{t} := \epsilon \ket{\bar{t}}, \quad \mbox{with } \epsilon := \fourmat{0}{-1}{1}{0}.
\eeq
Useful formulas are 
$$
[z|w]=\la w | z\ra,\qquad [z|w\ra = - [w|z\ra,
\qquad
[z|\vsigma|w]=-\la w |\vsigma| z\ra,\qquad [z|\vsigma|w\ra = [w|\vsigma|z\ra.
$$
} 
\be
\psi = \twovec{\ket{t}}{\ket{u}},
\ee
equipped with a norm $(\tl \psi,\psi) \equiv \bra{\tl u} t\ra + \bra{\tl t}u\ra$ 
and with Poisson brackets \cite{penrose_books}
\be \label{PBtu}
\{t^A , \bar{u}^B\} = -i\delta^{AB}, \qquad A,B = 0,1.
\ee
In \cite{twisted2}, it was shown that under symplectic reduction by the constraint Re$\braaket{u}{t}=0$, which means that the twistor has zero helicity, i.e. it is null, the phase space reduces to $T^*\SU(2)$ with its canonical Poisson algebra.
To generalize this construction to the Lorentz group, we double the set-up and start with two twistors, $\psi$ and $\tl \psi$. More precisely, we consider  
$\T_*=\T \setminus \{\bra{u}t\ra=0 \}$,
and define
\be
\T_2 \equiv \T_*\times\T_*.
\ee
This space with two copies of the Poisson brackets \Ref{PBtu}, is the space of our main interest.
Each space $\T$ separately carries a Hamiltonian representation of the $\sl(2,\C)$ algebra as quadratic functions of the spinors \cite{penrose_books}, a fact which was exploited in \cite{twisted3} to study the algebra of intertwiners.
To understand this representation and extend it to the cotangent bundle $T^*\slc$, it is useful to focus on the bi-spinor nature of the twistor. Therefore, for the moment, let us focus on one single copy of the twistor space $\T$, say the first copy in $\T_2 \equiv \T_*\times\T_*$.
From an algebraic viewpoint, the twistor represents a classical version of a Dirac bi-spinor, in the sense that $\psi$ is formed by two spinors, $\ket{t}$ right-handed and $\ket{u}$ left-handed.
The Hamiltonian representation of $\sl(2,\C)$ can then be defined by projecting\footnote{See appendix \ref{spinorial_appendix} for a standard definition of the abstract generators. There we also summarize our notations and conventions.} the generators $\cJ^{IJ}$ of the abstract $\sl(2,\C)$ algebra onto phase space generators $J^{IJ}$ via
\be
\{  J^{IJ}, \psi \} := i \cJ^{IJ} \psi.
\ee
One can easily check using \Ref{PBtu} that 
\be 
J^{IJ} = -\bar{\psi} \cJ^{IJ} \psi,
\ee
with $\bar{\psi} := \psi^\dagger \gamma^0=(-\bra{u},-\bra{t})$, form a Poisson representation of the Lorentz algebra 
\be
[ \cJ^{IJ}, \cJ^{KL} ] = -i \left(  \eta^{JK}\cJ^{IL} + \eta^{IL}\cJ^{JK} - \eta^{IK}\cJ^{JL} - \eta^{JL} \cJ^{IK} \right).
\ee
More specifically, the rotation and boost generators $J_i := \f12 \eps_{ijk} J^{jk}$ and $K_i:=J^{0i}$ are given by
\be\label{defJK}
\vec{J} = {\rm Re} \braaaket{t}{\vec{\sigma}}{u}, \qquad \vec{K} = {\rm Im} \braaaket{t}{\vec{\sigma}}{u},
\ee
and the right-left $\su(2)$ generators $\vJ^{L,R} := (\vJ\pm i \vK)/2$ by
\be \label{defJLJR}
\vJ^L=\f12\la t|\vsigma|u\ra,
\qquad
\vJ^R=\f12\la u|\vsigma|t\ra.
\ee
The action of these generators on the spinors can be easily calculated from \Ref{PBtu},
\be\label{poisson_t_u}
\{ \vec{J}^L , \ket{u} \} = \frac{i}{2} \vec{\sigma} \ket{u}, \qquad \{ \vec{J}^R , \ket{t}\} = \frac{i}{2} \vec{\sigma} \ket{t},
\ee
which exponentiates to an $\SL(2,\C)$ action in the defining representation on the spinors,
\be\label{SL2C_action1}
\ket{u} \rightarrow e^{\frac{i}{2} \vec{p}_L \cdot \vs } \ket{u},
\qquad \ket{t} \rightarrow e^{\frac{i}{2} \vec{p}_R \cdot \vs } \ket{t},
\ee
where $\vec{p}_L = \overline{\vec{p}_R}$ are a complex coordinate system on the left- and right-handed copy of 
$\SU(2)$ respectively.\footnote{Notice that the action is flipped on the dual spinors,
\be
\{ \vec{J}^L , \ketr{t} \} = \frac{i}{2} \vec{\sigma} \ketr{t}, \qquad
\{ \vec{J}^R , \ketr{u} \} = \frac{i}{2} \vec{\sigma} \ketr{u}, \qquad
\ketr{u} \rightarrow e^{\frac{1}{2} \vec{p}_R \cdot \vs    } \ketr{u}, \qquad
\ketr{t} \rightarrow e^{\frac{1}{2} \vec{p}_L \cdot \vs    } \ketr{t}.
\ee
}

Finally, the familiar quadratic scalars in $\psi$ are encoded in the single complex invariant $\braaket{u}{t}$,
\be
\bar\psi \psi = -2\, {\rm Re}\, \bra{u}{t}\ra, \qquad \bar\psi\g^5 \psi = -2i\, {\rm Im}\, \braaket{u}{t},
\ee
or in terms of the Casimirs,
\be\label{Cas}
J^2-K^2 = ({\rm Re}\, \bra{u}{t}\ra)^2 - ({\rm Im}\, \bra{u}{t}\ra)^2,
\qquad \vJ\cdot\vK = -{\rm Re}\, \bra{u}{t}\ra \, {\rm Im}\, \bra{u}{t}\ra.
\ee

Before moving on, let us comment on the ``non-diagonal'' form of the symplectic structure \Ref{PBtu}. This can be traced back to the invariance of the norm in twistor space under SU(2,2), the covering group of the conformal group.
For some purposes, it is useful to consider also a rotated pair of spinors which diagonalizes the symplectic structure. 
This can be achieved by the following linear combinations,
\be
\ket{z} := \frac{1}{\sqrt{2}}\Big(\ket{t} + \ket{u}\Big),  \qquad \ketr{w} := \frac{i}{\sqrt{2}}\Big(\ket{t} - \ket{u}\Big),
\ee
which diagonalize the symplectic structure as
\be \label{symplectic_z_w}
\{  z^A, \bar{z}^B \} = -i \delta^{AB}, \qquad \{  w^A, \bar{w}^B \} = -i \delta^{AB},
\ee
with remaining Poisson brackets vanishing.\footnote{The diagonal Poisson brackets are however not $\slc$-invariant, since the boosts mix $\ket{z}$ and $\ket{w}$.}
These spinors are used as a starting point in \cite{spinor3},\footnote{The definition used here differs by a factor $\frac{1}{\sqrt{2}}$ from the one used in \cite{spinor3}. This choice is more convenient in this context since it corresponds to a unitary transformation on the space of $\gamma$-matrices (see appendix \ref{Dirac_appendix}).}
and are essentially a double copy of the spinor variables used in the twisted geometries approach for $\SU(2)$ \cite{twisted1, twisted2}, therefore they carry the same interpretation. In particular, their norm can be interpreted as the area of a face of a polyhedral decomposition of the spatial manifold. Depending on the context, it will be advantageous to work with either the 'chiral' variables $\ket{t}, \ket{u}$ or the 'real' variables $\ket{z}, \ket{w}$. 

In terms of the $\ket{z}, \ket{w}$, the $\sl(2,\C)$ generators take the following form,
\be
\vJ=\f12(\la z|\vsigma|z\ra+\la w|\vsigma|w\ra),
\qquad
\vK=\f12([w|\vsigma|z\ra+\la z|\vsigma|w])\, ,
\ee
and
\be\label{utAiB}
\la u|t\ra= \frac{1}{2}(A-iB),
\ee
with real Poisson invariants
\be
A = \braaket{z}{z} - \braaket{w}{w},
\qquad
B = \braakettr{z}{w} + \braaketrt{w}{z}. \label{AB_zw}
\ee
Thus we have represented $\sl(2,\C)$ on $\T$ with phase space
functions $\vec{J}_L, \vec{J}_R$. Alternatively, we could have represented $\sl(2,\C)$ in the
second copy $\T$ in $\T_2$: this other choice is denoted with tilded variables
 $\vec{\tilde{J}}_L, \vec{\tilde{J}}_R$. These two choices correspond to chosing left- and rightinvariant vectorfields on $\SL(2,\C)$ as generators of its Lie-algebra respectively.

\subsection{Phase space reduction and holonomy-flux algebra for $\SL(2,\C)$ }

Thus far, we have represented the $\sl(2,\C)$ algebra on the space $\T$. 
To obtain the cotangent bundle $T^*\slc\cong \sl(2,\C)\times \slc$, we need also a representation of group elements. These can no longer be represented on a single copy $\T$ of twistor space, for this purpose the full space $\T_2$ is needed.
A map $(\ket{t},\ket{u},\ket{\tl t},\ket{\tl u})\mapsto G\in \slc$ can be easily obtained if we take, for all spinors, the unique group element $G$ mapping one pair $(\ket{\tl t}, \ket{\tl u})$ into the other $(\ket{t}, \ket{u})$. This is given by\footnote{When using the spinors $\ket{z}, \ket{w}$ which diagonalize the symplectic structure instead, the $\SL(2,\C)$ element takes the following form:
\be
G = \f1{\sqrt{(A-iB)(\tl A-i\tl B)}}
\bigg[ | z \ra [ \tz | - | z ] \la \tz |  - \left( | w \ra [ \tilde{w} | - | w ] \la \tilde{w} | \right)  + i \left( | z \ra \la \tilde{w} | + | z ][ \tilde{w} |\right) - i \left( | w \ra \la \tz | + | w ][ \tz |\right) \bigg]
\ee
Up to the normalization, it is sum of four $\SU(2)$ elements. 
Of course, one could parameterize $G$ in a different way, to be a simple expression in $\ket{z}, \ket{w}$, however the price to pay is that the denominator is in general not an $\sl(2,\C)$ invariant. From this point of view the chiral variables are preferred.
}
\be \label{Group_element}
G := \Group{t}{\tt}{u}{\tu},
\ee
which satisfies
\be  \label{sl2c_on_spinors}
G \frac{\ket{\tt}}{\sqrt{\braaket{\tu}{\tt}}} = -\frac{\ketr{u}}{\sqrt{\braaket{u}{t}}}, \qquad G \frac{\ketr{\tu}}{\sqrt{\braaket{\tu}{\tt}}} = \frac{\ket{t}}{\sqrt{\braaket{u}{t}}}.
\ee
Using the identity $2\det M=(\tr M)^2-\tr M^2$ valid for all 2$\times$2 matrices $M$, it is straightforward to check that $\det G = 1$ and thus $G \in \SL(2,\C)$. 
Notice that $G$ is by construction fully right-handed. A left-handed group element can be obtained via the dual map \Ref{dual}, or by hermitian conjugation,
\be
G^\dagger = \frac{\ketr{\tt}\bra{t}-\ket{\tu}[u|}{\sqrt{\braaket{t}{u}\braaket{\tt}{\tu}}}.
\ee

The group elements obtained in this way are well defined in $\T_2$, where the restrictions $\bra{u}t\ra\neq 0$, $\bra{\tl u}\tl t\ra\neq 0$ apply. To map $\T_2$ into $T^*\slc$, we consider a similar restriction in the target space, and define
$T^*\slc_* \equiv T^*\slc \setminus \{|J|=0\}$. Then, the quantities \Ref{defJLJR} (or equivalently \Ref{defJK}) and \Ref{Group_element} are our candidate coordinates for the submanifold $T^*\slc_*$ in $\T_2$.
Consider now the complex constraint
\be\label{defM}
\cM := \braaket{u}{t} - \braaket{\tu}{\tt}.
\ee
This constraint imposes the matching of the ``complex helicities'' of the twistors,\footnote{The real helicity of a twistor is defined as Re$\bra{u}t\ra$.} and generates a $\U(1)^\C \simeq \C$ action in phase space,
\beq\label{Maction}
\ket{t} \mapsto e^{+\frac{i}{2}\beta}\ket{t}, \qquad \ket{u} \mapsto e^{+\frac{i}{2}\bar{\beta}} \ket{u}, \qquad
\ket{\tt} \mapsto e^{-\frac{i}{2}\beta}\ket{\tt}, \qquad \ket{\tu} \mapsto e^{-\frac{i}{2}\bar{\beta}}\ket{\tu}, \qquad \beta\in\C.
\eeq
By analogy with the SU(2) case, we will refer to \Ref{defM} as the \emph{area matching constraint}.
Symplectic reduction by this constraint eliminates four dimensions in phase space, thus the initial 16-dimensional space $\T_2$ is reduced to a 12-dimensional space. We are now ready to state the following:
{\prop{
The symplectic reduction of $\T_2$ by $\cM$ gives the cotangent bundle of the Lorentz group with its canonical Poisson algebra, 
\be
\T_2 /\!/ U(1)^\C \cong T^*\slc_*.
\ee
}}
\begin{proof}
To prove our claim, we first check that the coordinates \Ref{defJLJR} and \Ref{Group_element} Poisson-commute with the constraint, then compute the induced Poisson algebra.
The first step is trivial: As stated in (\ref{Cas}) the constraint $\cM = (A-\tilde{A}) - i (B - \tilde{B})$ decomposes into the Casimir invariants of $\sl(2,\C)$, hence
\be
\{ \cM, \vec{J} \} = \{ \cM, \vec{K} \} = 0.
\ee
Then, an explicit calculation gives
\be
\{ \cM, G \} = 0.
\ee
The second step requires the evaluation of the Poisson brackets among the generators (both, the tilded and untilded ones) and the group element. This is a simple, if lengthy, calculation, which gives
\begin{align}
&\{ J^L{}_i, J^L{}_j \}  =   \epsilon^{ijk} J^L{}_k, &\{ J^R{}_i, J^R{}_j \} =  \epsilon^{ijk} J^R{}_k, 
&& \{ J^L{}_i , J^R{}_j\} = 0  &\, \\
&\{\tl J^L{}_i, \tl J^L{}_j \}  =   \epsilon^{ijk} \tl J^L{}_k, &\{ \tl J^R{}_i, \tl J^R{}_j \} =  \epsilon^{ijk} \tl J^R{}_k, 
&& \{\tl J^L{}_i , \tl J^R{}_j\} = 0 &\, \nn
\end{align}
\begin{align}
& \{ G^A{}_B, G^C{}_D \}  =  0 \nn
\end{align}
\begin{align}
&\{\vec{J}^R , G \}  = \f{i}2\vec\sigma G, &\{\vec{\tl J}{}^R , G \}  = -\f{i}2G \vec\sigma, 
&&\{ \vec{J}^L, G \}  =  0, &&\{ \vec{\tl J}{}^L, G \}  =  0, & \nn \\
&\{\vec{J}^L, G^\dagger \}  = - \f{i}2G^\dagger\vec\sigma, 
&\{\vec{\tl J}{}^L, G^\dagger \}  = \f{i}2\vec\sigma G^\dagger, 
&&\{ \vec{J}^R, G^\dagger \}  =  0, &&\{ \vec{\tl J}{}^R, G \}  =  0. \nn
\end{align}
This is the expected Poisson algebra for $T^*\SL(2,\C)$ with a group element $G$ in the defining right-handed representation 
$\bf{(0,1/2)}$. By taking the hermitian conjugate $G^\dagger$, or alternatively by exchanging the spinors for their duals and vice versa in (\ref{Group_element}), one gets a left-handed representation $\bf{(1/2,0)}$.
\end{proof}
As it turns out, the brackets computed above are actually valid also off the ``mass-shell'' of \Ref{defM}, except for the one with two group elements. For this to vanish, one has to be on the constraint surface $\cM = 0$.

\subsection{Properties of the group element}

First of all, notice that on the constraint surface the normalizations  in \Ref{sl2c_on_spinors} cancel each other. Therefore, on-shell the group element maps the spinors exactly,
\be  \label{Gtu}
G \ket{\tt} = -\ketr{u}, \qquad G \ketr{\tu} = \ket{t}, \qquad
G^\dagger \ketr{t} = -\ket{\tu}, \qquad G^\dagger \ket{u} = \ketr{\tt}.
\ee
This transformation property can be translated at the level of the bivectors. Seeing them as $2\times 2$ matrices via the map
$J\:=\vJ \cdot \vec\s$, we have
\be\label{JGJG}
\Big(\tl J^R, \tl J^L\Big) = \Big(-G^{-1}J^R G, -G^\dagger J^L (G^\dagger)^{-1}\Big)
\ee
Hence, the group element $G$ parallel transports the bivector $J$ into $\tl J$, and can be interpreted as the holonomy of an $\SL(2,\C)$ connection along an edge with $J$ on its source vertex and $\tl J$ on its target. 

As a function parameterizing a 6-dimensional space in 16-dimensional $\T_2$, \Ref{Group_element} has a 10-dimensional group of isometries.
Six of this are the invariance under a 'twisted' action of the group, like the one for the $\SU(2)$ case found in \cite{spinor2}. This acts simultaneously on the spinors in both reference frames as
\beq
\ket{t} \rightarrow \Lambda \ket{t}, &\qquad & \ket{u} \rightarrow (\Lambda^{-1})^\dagger \ket{u}, \nn \\
\ket{\tt} \rightarrow G^{-1} \Lambda G \ket{\tt}, & \qquad & \ket{\tu} \rightarrow G^\dagger (\Lambda^{-1})^\dagger (G^{-1})^\dagger \ket{\tu},
\eeq
where $\Lambda$ is an arbitrary matrix of $\SL(2,\C)$ in the defining representation. This 'twisted' rotation has a natural interpretation: As the group element $G$ measures how the local reference frames at the initial and final vertex of an edge are rotated \emph{with respect to each other}, a simultaneous rotation in both frames does not affect $G$. However, in order to make such a translation meaningful, the spinors at the final vertex first have to be parallel transported to the initial vertex, then can be rotated, and then have to be parallel transported back. This explains the appearance of the group element $G$ (which defines the parallel transport) in the rotations for the 'tilded' variables.

The remaining are the following four real rescalings,
\beq
&& \twovec{\ket{t}}{\ket{u}} \rightarrow \twovec{a\ket{t}}{a\ket{u}}, \qquad a\in \R^+, \qquad
\twovec{\ket{\tt}}{\ket{\tu}} \rightarrow \twovec{b\ket{\tt}}{b\ket{\tu}}, \qquad b\in \R^+, \nn \\
&& \twovec{\ket{t}}{\ket{\tt}} \rightarrow \twovec{c\ket{t}}{{c^{-1}}\ket{\tt}}, \qquad c\in \R^+, \qquad
\twovec{\ket{u}}{\ket{\tu}} \rightarrow \twovec{d\ket{u}}{{d^{-1}}\ket{\tu}}, \qquad d\in \R^+.
\eeq

\section{Covariant Twisted Geometries}

\label{cov_twist}
A beautiful aspect of the spinorial variables for loop quantum gravity is to admit a simple geometric interpretation, given by twisted geometries \cite{twisted1,twisted2,twisted3,polyhedron}. These are a collection of polyhedra associated with a cellular decomposition dual to the graph, described by 3-dimensional area vectors (the spinor vectors $X=\bra{z}\f\sigma2\ket{z}$), and angles representing the extrinsic curvature among them (the spinor phases).
This construction extends to the $\slc$ case, where it is related to spacelike Lorentzian polyhedra when the simplicity constraints hold \cite{spinor3}. To appreciate the covariant version of twisted geometries, one needs a decomposition of the $\slc$ group element on each edge in terms of bivectors, representing the 4-dimensional area normals, and an angle characterizing again the extrinsic curvature.

Let us first recall the SU(2) case. There $(X,g)\in T^*\SU(2)$ with $\tl X=-g^{-1}Xg$, analogously to \Ref{JGJG}, and the group element can be further parameterized, in the fundamental representation, as \cite{twisted1,twisted2}
\be\label{defg}
g=n(\zeta)e^{-\frac{i}{2}\xi\s_3}\eps^{-1}\tl n^{-1}(\tl\zeta),
\ee
where 
\be
n(\zeta)=\f{1}{\sqrt{1+|\zeta|^2}}\,\mat{cc}{1 & \zeta \\ -\bzeta & 1}
\ee
is the Hopf section of the coset $S^2 = \SU(2)/\U(1)$. The presence of $\eps$ in \Ref{defg} flips the orientation of the normals $X$ and $\tl X$ and allows to preserve the same sign in their Poisson brackets. 
The parameterization can be also written as $g=n(\zeta)e^{-\frac{i}{2}\phi\s_3}\eps^{-1}(\tl n(\tl\zeta) e^{-\frac{i}{2}\tl\phi\s_3})^{-1},$ with $\xi\equiv \phi-\tl\phi$, where 
\be
n(\zeta)e^{-\frac{i}{2}\phi\s_3} \twovec{1}{0} := \frac{-\ketr{z}}{\sqrt{\braaket{z}{z}}}, \qquad
n(\zeta)e^{-\frac{i}{2}\phi\sigma_3} \twovec{0}{1} := \frac{\ket{z}}{\sqrt{\braaket{z}{z}}}.
\ee

This construction can be easily generalized to \Ref{Group_element}.
First, we factorize $G=G(t,u)\eps^{-1}G(\tt,\tu)^{-1}$,
with
\be\label{pluto}
G(t,u) \twovec{1}{0} := \frac{-\ketr{u}}{\sqrt{\braaket{u}{t}}}, \qquad
G(t,u) \twovec{0}{1} := \frac{\ket{t}}{\sqrt{\braaket{u}{t}}},
\qquad G(t,u) := \frac{1}{\sqrt{\braaket{u}{t}}} \fourmat{\bar{u}^1}{t^0}{-\bar{u}^0}{t^1}.
\ee
Then, we decompose this group element following the Iwasawa decomposition,
\be \label{decomposition}
G=n(\zeta)\,T_\alpha\,e^{-\frac{i}{2}\Phi\s_3},\qquad
(\zeta,\alpha,\Phi)\in\C^3,
\ee
where $n(\zeta)$ is the same Hopf section defined above, and 
\be
T_\alpha \,=\, \mat{cc}{1 & \alpha\\ 0 & 1},
\ee
is the subgroup of upper triangular matrices in $\SL(2,\C)$. They are the block allowing to go from an orthogonal basis to a non-orthogonal basis of the spinor space.

We easily compute
\be
n(\zeta)\,T_\alpha\,e^{-\frac{i}{2}\Phi\s_3} \,\mat{c}{1\\0}\,=\,
\f{e^{-\frac{i}{2}\Phi}}{\sqrt{1+|\zeta|^2}}\mat{c}{1\\ -\bzeta}
,\qquad
n(\zeta)\,T_\alpha\,e^{-\frac{i}{2}\Phi\s_3} \,\mat{c}{0\\1} \,=\,
\f{e^{\f{i}2\Phi}}{\sqrt{1+|\zeta|^2}}\mat{c}{\alpha+\zeta\\ 1-\alpha\bzeta}.
\ee
Comparing these with \Ref{pluto}, we determine the parameters $\zeta, \alpha, \Phi$ as
\be
\zeta=\frac{{u}^0}{{u}^1},
\qquad
{\rm Re} \,\Phi = 2\arg u^1, \qquad {\rm Im}\,\Phi = \ln\f{\bra{u}u\ra}{\bra{u}t\ra}, \qquad
\alpha=- \exp\{-2i\arg u^1\} \f{[t| u\ra}{\la u|t\ra}.
\ee
The first two can be recognized as the same decomposition of the SU(2) case \cite{twisted2}, whereas the third and fourth capture the boost dependence on the spinors.

Putting together the two parts of the $\SL(2,\C)$ holonomy along the edge, we obtain the covariant twisted geometry decomposition of the $\SL(2,\C)$ group element in a form similar to \Ref{defg},
\be\label{ganzo}
G(t,u,\tt,\tu) = G(t,u)\eps^{-1}G(\tt,\tu)^{-1}
\,=\, n(\zeta)\,T_\alpha\,e^{-\frac{i}{2}(\Phi-\tl\Phi)\s_3} \bar T_{\tl\alpha} \eps^{-1} \tl n^{-1}(\tl\zeta),
\ee
where 
\be
\bar T_\alpha \,=\, \mat{cc}{1 & 0 \\ \alpha & 1}.
\ee
In particular, $\Xi={\rm Re}\,\Phi-{\rm Re}\,\tl\Phi$ is the quantity carrying information on the 4-dimensional dihedral angle.

The SU(2) case can be immediately obtained setting $\ket{t} = \ket{u}$, which implies 
Im $\Phi= \alpha=0$, so that \Ref{ganzo} reduces to \Ref{defg}. 

This classical decomposition of the $\SL(2,\C)$ holonomy will be helpful for the geometric interpretation of covariant twisted geometries, and in building appropriate coherent states and spin foam amplitudes for the Lorentzian theory.
We postpone a complete study of covariant twisted geometries and their Poisson brackets to future work.

\section{Twistor networks}

\subsection{Generalizing Spinor Networks to $\SL(2,\C)$}
Up to now, we have discussed a single copy of the $\sl(2,\C)$ algebra. From the perspective of loop quantum gravity and spinfoams, we would like to be able to describe spin network states for $\SL(2,\C).$\footnote{See for example \cite{freidel_livine_spinnets, barrett_lorentzian} for a rigorous definition of spin networks for the non-compact group $\SL(2,\C)$.} For this purpose, we choose an arbitrary oriented graph $\Gamma$, closed and connected for simplicity, and we attach one copy of $T^*\slc$ to each edge of the graph. Upon describing $T^*\slc$ in terms of $\T_2$, we obtain a notion of \emph{twistor network}: a graph labeled by two twistors in $\T_2$ per edge, or equivalently one twistor per half-edge. 
This generalizes the spinor networks for SU(2) studied in \cite{spinor2} (see also \cite{spinor1,holosimplicity}), which carry one twistor per edge, or equivalently a spinor per half-edge. That is, in going from SU(2) to $\slc$, we simply double the number of twistors.

To keep the notation simple, we use the orientation of each edge to uniquely identify its source and target vertices $s$ and $t$, and eliminate the tildes defining $\ket{t}=\ket{t^s}, \ket{\tt}=\ket{t^{t}}$.
We then have a pair of twistors or bi-spinors $\ket{t^v_e}, \ket{u^v_e}$ on each edge, carrying a representation of $\sl(2,\C)$ associated with the invariant $\braaket{u_e^v}{t_e^v}$. To have a unique invariant per edge, we impose the matching conditions
\be \label{defMe}
\cM_e := \braaket{u^s_e}{t^s_e} - \braaket{u^t_e}{u^t_e} \stackrel{!}{=} 0,
\ee
that is
\be
{\rm Re}\cM_e \equiv A_e^s - A_e^t \stackrel{!}{=} 0, \qquad {\rm Im}\cM_e \equiv B^t_e - B^s_e \stackrel{!}{=} 0.
\ee
They ensure that the two Casimirs of the $\sl(2,\C)$ representations living at the source and target vertices coincide, and so do the two representations.
The constraints generate $\U(1)^\C \sim \C$ transformations on the spinors $\ket{t}$ and $\ket{u}$,
\be \label{action_of_h}
|t^s_e\ra\arr e^{+i\beta_e}|t^s_e\ra,
\quad
|t^t_e\ra\arr e^{-i\beta_e}|t^t_e\ra,
\quad
|u^s_e\ra\arr e^{+i\bar{\beta}_e} |u^s_e\ra,
\quad
|u^t_e\ra\arr e^{-i\bar{\beta}_e} |u^t_e\ra, \qquad \beta\in\C,
\ee
that is, simultaneous rescalings.

Furthermore, we require invariance under global $\SL(2,\C)$ transformations at each vertex $v$ of the graph. This is imposed by the $\SL(2,\C)$ closure constraints
\be
\label{closure}
\cC_v
\,\equiv\,
\sum_{e\ni v} \la t^v_e |\vsigma|u^v_e\ra =0,
\ee
corresponding to its real and imaginary parts,
\be
\label{closure_real}
\cC_v^J
\equiv 
\sum_{e\ni v} \vJ^v_e = 0,
\qquad 
\cC_v^K
\equiv
\sum_{e\ni v} \vK^v_e
=0. \nn
\ee
From these expression, it is obvious that $\cC_v^J$ and $\cC_v^K$
generate global $\SL(2,\C)$ transformations on all the twistors $(\ket{t^v_e},\ket{u^v_e})$ attached the vertex $v$.
The geometrical interpretation in terms of 3d polyhedra in Minkowski spacetime is discussed in \cite{spinor3}. Here we would like to introduce the action principle summarizing the phase space structure with its constraints, on a given graph $\Gamma$:
\bes\label{action}
S_\Gamma[t^v_e,u^v_e]
&\equiv \,\int d\tau\,&\Big(
\sum_e -i\la u_e^{s,t}|\pp_t t_e^{s,t}\ra -i\la t_e^{s,t}|\pp_t u_e^{s,t}\ra\\\nn
&&+\sum_e \Phi_e (\la u_e^s|t_e^s\ra-\la u_e^t|t_e^t\ra)
+\sum_{v}\sum_{e\ni v} \la t^v_e|\Theta_v|u^v_e\ra\Big).
\ees
The kinetic term encodes the canonical Poisson bracket, for which $\ket{t}$ is canonically conjugate to $\bra{u}$ and vice-versa. All the constraints are first class. The complex Lagrange multiplier $\Phi_e$ imposes the complex area matching constraints \Ref{defMe}, and 
the complex traceless matrix $\Theta_v$ is the Lagrange multiplier enforcing the $\SL(2,\C)$-closure constraints at each vertex. 

To summarize, a twistor network is the generalization to the Lorentzian case of a spinor network, that is a set of twistors or bi-spinors $\ket{t_e^{s,t}}, \ket{u_e^{s,t}}$ 
satisfying both the matching and closure constraints, and up to the corresponding $\C^E$ and $\SL(2,\C)^V$ transformations. These are thus elements of the symplectic quotient $\C^{8E}/\!/(\C^E\times\SL(2,\C)^V)$, a phase space of dimensions $2\times 6(E-V)$, and which is isomorphic to the phase space over the configuration $\SL(2,\C)^E/\SL(2,\C)^V$ corresponding to spin networks on the graph $\Gamma$ for the gauge group $\SL(2,\C)$.

The structure of constraints of a twistor network can be represented in the following scheme:

\begin{center}
\begin{tabular}{ccc}

{Twistor space} $\underset{e}{\times} \T_2$ \qquad & $\longrightarrow$ & \ holonomy-flux phase space $\underset{e}{\times} T^*\slc$ \\
& \emph{matching area} & \\
$\downarrow$ \emph{closure} & & $\downarrow$ \emph{closure} \\
& & \\
$\GL(N,\C)$ formalism \qquad & $\longrightarrow$ & \ gauge-invariant phase space  \\
& \emph{matching area} & 
\end{tabular}
\end{center}

The usual path of constraints implementation is right-bottom: one first imposes the area matching to reduce the twistor structure to the standard holonomy-flux algebra, then imposes gauge invariance at the vertices. However, one can proceed otherwise, and impose first the closure constraint.
This alternative is very interesting, because it introduces a simple set of $SL(2,\C)$ observables. These have been studied in \cite{spinor3}, and contain a $\GL(N,\C)$ subalgebra. This is analogue of the SU(2) case, where working at the level of spinors one can characterize the algebra of SU(2) invariants in terms of a U(N) algebra. This framework has proved useful to address a number of questions \cite{un1,un2,un3}, and we believe the $\GL(N,\C)$ framework for the covariant case to be as prolific.

For completeness, we also give the corresponding expressions for the spinors $(\ket{z},\ket{w})$ with symplectic structure (\ref{symplectic_z_w}). The real and imaginary parts of the area matching constraints read
\be
\cM_e^A
\,:=\,
A_e^s-A_e^t
\,\equiv\,
(\la z^s_e|z^s_e\ra-\la w^s_e|w^s_e\ra)
-
(\la z^t_e|z^t_e\ra-\la w^t_e|w^t_e\ra)
\,\stackrel{!}{=}\,
0,
\ee
\be
\cM_e^B
\,:=\,
B_e^s-B_e^t
\,\equiv\,
(\la z^s_e|w^s_e]+[ w^s_e|z^s_e\ra)
-
(\la z^t_e|w^t_e]+[ w^t_e|z^t_e\ra)
\,\stackrel{!}{=}\,
0\, ,
\ee
and generate respectively $\U(1)$ transformations,
\be
|z^s_e\ra\arr e^{+i\theta_e}|z^s_e\ra,
\quad
|z^t_e\ra\arr e^{+i\theta_e}|z^t_e\ra,
\quad
|w^s_e\ra\arr e^{-i\theta_e} |w^s_e\ra,
\quad
|w^t_e\ra\arr e^{-i\theta_e} |w^t_e\ra,
\ee
and $\R$ transformations,
\bes
&&|z^s_e\ra\arr \cosh\eta_e|z^s_e\ra+ i\sinh\eta_e |w^s_e],
\quad
|z^t_e\ra\arr \cosh\eta_e|z^t_e\ra- i\sinh\eta_e |w^t_e],\nn\\
&&|w^s_e]\arr\cosh\eta_e|w^s_e]- i\sinh\eta_e|z^s_e\ra,
\quad
|w^t_e]\arr\cosh\eta_e|w^t_e]+ i\sinh\eta_e|z^t_e\ra.
\ees
The $\SL(2,\C)$-closure constraints become
\be
\cC_v^J
\,=\,
\sum_{e\ni v} \vJ^v_e
\,=\,
\f12\sum_{e\ni v} \la z^v_e|\vsigma |z^v_e\ra +\la w^v_e|\vsigma |w^v_e\ra
\,=\,
0\,,
\ee
\be
\cC_v^K
\,=\,
\sum_{e\ni v} \vK^v_e
\,=\,
\f12\sum_{e\ni v} [w^v_e|\vsigma |z^v_e\ra +\la z^v_e|\vsigma |w^v_e]
\,=\,
0\,.
\ee
Finally, the action principle on a given graph is given by
\bes
S_\Gamma[z^v_e,w^v_e]
&\equiv \,\int d\tau\,&\Big(
\sum_e -i\la z_e^{s,t}|\pp_t z_e^{s,t}\ra -i\la w_e^{s,t}|\pp_t w_e^{s,t}\ra
+\sum_e \phi_e (A_e^s-A_e^t) +\tphi_e (B_e^s-B_e^t) \nn\\
&&+\sum_{v}\sum_{e\ni v} \la z^v_e|\Theta_v|z^v_e\ra +\la w^v_e|\Theta_v |w^v_e\ra
+[w^v_e|\tilde{\Theta}_v|z^v_e\ra +\la z^v_e|\tilde{\Theta}_v |w^v_e]\Big)
\ees
where $\phi_e,\tphi_e\in\R$ are the Lagrange multipliers enforcing the matching constraints $\cM_e^A$ and $\cM_e^B$ on each edge $e$, while the traceless Hermitian matrices $\Theta_v$ and $\tilde{\Theta}_v$ enforce the closure constraints $\cC_v^J$ and $\cC_v^K$ at each vertex $v$.

\subsection{Action principle: from Twistors to Group Elements}

The action $S[t^{s,t}_e,u^{s,t}_e]$ can be also given a ``first order'' formulation, in terms of both spinors and holonomies.
To do so, we insert
\be\label{pippo}
G_e\,|t^t_e\ra\,=\,-|u^s_e], \qquad G_e\,|u^t_e]\,=\,|t^s_e\ra
\ee 
on each edge in the kinetic terms of \Ref{action}. After a few algebraic manipulations (using that \linebreak $G^{-1}=\eps\,G^T\,\eps^{-1}$ for all $\SL(2,\C)$ group elements), one gets
\be
\int d\tau\,\Big(-i\la u_e^{s,t}|\pp_t t_e^{s,t}\ra -i\la t_e^{s,t}|\pp_t u_e^{s,t}\ra\Big)
\,=\,
\int d\tau\,
\Big(-i[t^t_e|G^{-1}\pp_t G|u^t_e]+i\la t^t_e|(\pp_t G^{-1})G|u^t_e\ra\Big),
\ee
where we have discarded a total derivative term. The two terms on the right-hand side are the complex conjugate of each other, and the derivative of the group element lies in the Lie algebra, $G^{-1}\pp_t G\in\sl(2,\C)$.
The area matching conditions are then traded for new constraints imposing \Ref{pippo}, so that overall, \Ref{action} can be written with spinors and holonomies as independent variables as
\bes\label{action1}
&& S_\Gamma[t^v_e,u^v_e,G_e]
=\int d\tau\,\Big(
\sum_e -i[t^t_e|G_e^{-1}\pp_t G_e|u^t_e]+i\la t^t_e|(\pp_t G_e^{-1})G_e|u^t_e\ra\\
&& \qquad +\sum_e
\la T_e|G_e|t^t_e\ra+\la T_e|u^s_e]
+\la U_e| G_e\,|u^t_e]-\la U_e|t^s_e\ra
+\sum_{v}\sum_{e\ni v} \la t^v_e|\Theta_v|u^v_e\ra\Big), \nn
\ees
where $T_e$ and $U_e$ are the new Lagrange multipliers.

\subsection{Reduction to $\SU(2)$ and Simple Twistor Networks}
\label{recoverSU2}
In this section we study the `$\SU(2)$-limit' of the $\SL(2,\C)$ elements defined so far.
This allows us to uncover in this new language a number of structures which appear in spin foam models.
What condition has to be imposed on the four spinors to make $G(t,u,\tt,\tu)$ unitary? It is easy to see that this is indeed the case if we set
\be
\ket{u} \stackrel{!}{=} \ket{t} ,
\qquad
\ket{\tu} \stackrel{!}{=} \ket{\tt}.
\ee
This condition corresponds to $\ket{w}\equiv 0$, thus all the 3d-geometric information is captured by the single spinor $\ket{z}$, as it is in the pure $\SU(2)$ case \cite{spinor1,spinor2}. Accordingly, we just denote $\ket{u} = \ket{t} =\ket{z}$, $\ket{\tu} = \ket{\tt} =\ket{\tz}$.

We then have $J^R(z,z)\equiv J^L(z,z)$ trivially, and the $\SL(2,\C)$ group element \Ref{Group_element} reduces to the spinorial form of the $\SU(2)$ holonomy given in \cite{twisted2} (see also \cite{spinor1,spinor2}),
\be
G(z,\tz)=\f{|z\ra[ \tz|-|z]\la\tz|}{\sqrt{\la z|z\ra \la \tz|\tz\ra}}
\quad\in\,\SU(2).
\ee
The mapping property \Ref{sl2c_on_spinors} still applies, now in the form
$$
G\,\f{|\tz\ra}{\sqrt{\la \tz|\tz\ra}}
\,=\,
\,-\f{|z]}{\sqrt{\la z|z\ra}},
\qquad
G\,\f{|\tz]}{\sqrt{\la \tz|\tz\ra}}
\,=\,
\,\f{|z\ra}{\sqrt{\la z|z\ra}}.
$$

Similarly, we can reduce the group element $G$ to any $\SU(2)$ subgroup of $\SL(2,\C)$, defined acting upon the canonical with a pure boost, $\SU(2)_\Lambda\,\equiv\,\Lambda\SU(2)\Lambda^{-1},$ 
$\Lambda=\Lambda^\dagger=e^{\vec{b}\cdot\vsigma} \in\SL(2,\C)$.
This defines the non-canonical embedding 
\be
G=\Lambda
\f{| z\ra[ \tz|-| z] \la \tz|}{\sqrt{\la  z| z\ra\la \tz|\tz\ra}}
\Lambda^{-1}
\quad\in\,\SU(2)_\Lambda\,.
\ee
This corresponds to a group element in the generic form \Ref{Group_element}, where the left and right spinors are obtained boosting the same initial spinor in opposite directions:
\be
\label{transition}
|t\ra=\Lambda| z\ra,
\qquad
|u\ra=\Lambda^{-1}| z\ra,
\qquad
\ket{\tt} = \Lambda \ket{\tilde{ z}},
\qquad
\ket{\tu} = \Lambda^{-1} \ket{\tilde{ z}} \, .
\ee
Namely, we reconstruct a bivector associated to the pair $(\ket{t},\ket{u})$, from a single spinor $\ket{z}$ and a pure boost $\Lambda\in\SL(2,\C)/\SU(2)$. The latter defines a time-normal (a unit future-oriented time-like 4-vector).
But this is precisely what happens when we impose the simplicity constraints in spin foam models: a simple bivector is entirely determined by a 3-dimensional vector, and a 4-dimensional timelike normal.

Generalizing this procedure from a single edge to a whole graph $\Gamma$, we introduce a notion of \emph{simple twistor networks}, as the special class of twistor networks satisfying the simplicity constraints, and thus entirely determined by a spinor network plus an assignment of boosts at the vertices.
In fact, as shown in \cite{spinor3}, requiring the holomorphic simplicity constraints provides one boost $\Lambda_v$ per vertex. We thus have two boosts on each edge, one at the source and one at the target vertex. In order to recover the $\SU(2)$ sector as above, we need to further require that the boosts are all the same at every vertex, i.e that the time-normals at all vertices are the same, or in other words require that the whole graph be interpreted as living in one (space-like) hypersurface. Such fixing is achieved by appropriate $\SL(2,\C)$ gauge transformations at every vertex and is just the time-gauge when we fix to $\Lambda_v=\id,\,\forall v$.

To see this in details, let us start with a spinor network on the graph $\Gamma$ defined in terms of spinors $ z^v_e\in\C^2$. The  action principle encoding the phase space structure of spinor networks on the graph $\Gamma$:
\be
S_\Gamma[ z^{s,t}_e\in\C^2]
\,\equiv\,
\int d\tau\,
\sum_e -i\la  z^{s,t}_e|\pp_t z^{s,t}_e\ra
+\sum_e \Phi_e(\la  z^{s}_e| z^{s}_e\ra-\la  z^{t}_e| z^{t}_e\ra)
+\sum_v \sum_{e\ni v} \la  z^v_e|\Theta_v| z^v_e\ra\,,
\ee
where $\Phi_e$ and $\Theta_v$ are Lagrange multipliers for the matching and closure constraints for $\SU(2)$.
Next, we perform arbitrary pure boosts at each vertex $\Lambda_v\in\SL(2,\C)$, $\Lambda_v^\dagger=\Lambda_v$,  as above in \eqref{transition}:
\be
\label{transition2}
|t^v_e\ra\equiv\Lambda_v| z^v_e\ra,
\qquad
|u^v_e\ra\equiv\Lambda_v^{-1}| z^v_e\ra\,.
\ee
It is straightforward to check that these define a set of twistors satisfying the matching and closure constraints for $\SL(2,\C)$ (as was already shown for the case of a single vertex in \cite{spinor3}):
\be
\forall e,\quad
\la u^s_e|t^s_e\ra
=\la  z^s_e|\Lambda_{s(e)}^{-1}\Lambda_{s(e)}| z^s_e\ra
=\la  z^s_e| z^s_e\ra
=\la  z^t_e| z^t_e\ra
=\la u^t_e|t^t_e\ra\,,
\ee
\be
\forall v,\quad
\sum_{e\ni v}|u^v_e\ra\la t^v_e|
=\Lambda_v\left(\sum_{e\ni v}| z^v_e\ra\la  z^v_e|\right)\Lambda_v^{-1}
\propto\id\,.
\ee
Considering equivalence classes of such sets of twistors under the action of $\SL(2,\C)$ transformations at every vertex and rescalings on every edge defines a simple twistor network. Notice that the boosts $\Lambda_v$ get actually re-absorbed in the $\SL(2,\C)$ gauge transformations acting at each vertex.

This defines a simple twistor network from a spinor network and boosts living at each vertex of the graph. We can also characterize directly the simple twistor networks without referring to the underlying spinor network. Indeed, let us consider the \emph{holomorphic simplicity constraints} introduced in \cite{spinor3},
given by
\be
\forall v,\quad
\forall e,f\ni v,\qquad
\cF^v_{ef}
\equiv
[t^v_e|t^v_f\ra-[u^v_e|u^v_f\ra=0.
\ee
As proved in \cite{spinor3}, these constraints imply that there exists an $\SL(2,\C)$ transformation, $\cG_v$, relating the right spinors $t^v_e$ to the left spinors $u^v_e$:
$$
\exists \cG_v\in\SL(2,\C),\quad
\forall e\ni v,\qquad
|t^v_e\ra=\cG_v\,|u^v_e\ra.
$$
Combining this to the $\SL(2,\C)$ closure constraints then implies the existence of the spinors $z^v_e$ and boosts $\Lambda_v$ such that \eqref{transition2} holds. 
This means that we can encapsulate the phase space structure underlying simple twistor networks in the following action principle: 
\bes\label{actionsimple}
&& S^{simple}_\Gamma[t^{s,t}_e,u^{s,t}_e]
=\int d\tau\,
\sum_e -i\la u_e^{s,t}|\pp_t t_e^{s,t}\ra -i\la t_e^{s,t}|\pp_t u_e^{s,t}\ra\\\nn
&& \qquad +\sum_e \Phi_e (\la u_e^s|t_e^s\ra-\la u_e^t|t_e^t\ra)
+\sum_{v}\sum_{e\ni v} \la t^v_e|\Theta_v|u^v_e\ra 
+\sum_v\sum_{e,f\ni v} \Psi_{e,f}([t^v_e|t^v_f\ra-[u^v_e|u^v_f\ra)),
\ees
where the new Lagrange multipliers $\Psi_{e,f}$ enforce the holomorphic simplicity constraints (for vanishing Immirzi parameter) $[t^v_e|t^v_f\ra=[u^v_e|u^v_f\ra$. This can easily be generalized to non-vanishing Immirzi parameter by introducing a non-trivial proportionality coefficient in the simplicity constraint, $[t^v_e|t^v_f\ra=e^{i\theta}[u^v_e|u^v_f\ra$, as shown in \cite{spinor3}.

This action principle represents a collection of 3d polyhedra with
arbitrary, Lorentzian extrinsic curvature among them. Notice that each
face has two bivectors associated to it, determining the 2-normal to
it in the reference frame of each polyhedron.
While the area of the bivectors match by virtue of the constraint, the
shape of the face will in general differ when reconstructed from one
polyhedron or the adjacent one. Hence, we have a discontinuous,
discrete geometry. Shape matching conditions are studied in \cite{BS}
for tetrahedra, and \cite{polyhedron} for arbitrary polyhedra. A 4d
action in which the shape matching conditions are imposed is studied
in \cite{Dittrich}, for Euclidean signature and $\Gamma$ dual to a
simplicial triangulation. It is interesting to compare our algebraic
construction of \Ref{actionsimple} with the purely geometric
construction of \cite{Dittrich}, something we hope to come back to in
future work.

These simple twistor networks are very interesting from the perspective that they contain the same information as a normal spinor network for $\SU(2)$, but allow to describe its natural embedding into a $\SL(2,\C)$-invariant structure, through the introduction of non-trivial time-normals living at each vertex of the graph $\Gamma$.
They provide a classical version of the simple projected spin networks \cite{lift}, which form the boundary Hilbert space of EPRL/FK spin foam models \cite{lift,covariance}.

\subsection{The Haar measure on $\SL(2,\C)$ in terms of spinors}

In this final section, we show that the formalism so far developed has another important application, because it allows us to rewrite integrals over the Lorentz group as complex integrals with simple measures. We achieve this by rewriting the Haar measure on $\slc$ in spinorial coordinates. In \cite{spinor2} we have shown that the Haar measure on $\SU(2)$ can be written as the product of two uncoupled Gaussian measures over the two spinors by directly checking the orthonormality of representation matrix elements of $\SU(2)$ with respect to that measure. Let us derive the same formula again, using a complementary method which directly generalizes to $\SL(2,\C)$.

We start with the $\SU(2)$-element $g$ written in terms of two spinors $z$ and $\tz$. Indeed, as shown in \cite{twisted2}, the group element
\be
g(z,\tz) := \frac{|z\ra [ \tz | - | z ] \la \tz |}{\sqrt{\braaket{z}{z} \braaket{\tz}{\tz}}}\,
\ee
is the unique $\SU(2)$ element mapping $z$ to $\tz$, or more explicitly satisfying:
\be
g\,\f{|\tz]}{\sqrt{\la \tz|\tz\ra}}\,=\,\f{|z\ra}{\sqrt{\la z|z\ra}},
\qquad
g\,\f{|\tz\ra}{\sqrt{\la \tz|\tz\ra}}\,=\,-\f{|z]}{\sqrt{\la z|z\ra}}\,.
\ee
This group element can always be decomposed as $g(z,\tz) = g(z)\eps^{-1}g^{-1}(\tz)$,
where the individual parts satisfy
\be
g(z)\,\mat{c}{0 \\1}
\,=\,\f{|z\ra}{\sqrt{\la z|z\ra}},
\qquad
g(z)\,\mat{c}{1 \\0}
\,=\,-\f{|z]}{\sqrt{\la z|z\ra}}, \qquad
g(z) = \frac{1}{\sqrt{\braaket{z}{z}}} \fourmat{\bz^1}{z^0}{-\bz^0}{z^1}, \label{su2_element_z}
\ee
and idem for $g(\tz)$.
Now we would like to write the integral $\int_{\SU(2)} dg f(g)$ in terms of $g=g(z,\tz)$ with integrations over the spinor variables.
By right-invariance of the Haar measure, we can always multiply the whole group element by $g(\tz)$ from the right without changing anything. Therefore it is enough to start with $g=g(z)$ and consider a group element which depends on only one spinor. This is due to the fact that the group element $g(z,\tz)$ carries information only about the \emph{relative} rotation of $\ket{z}$ and $\ket{\tz}$, but not about an absolute reference frame.
In terms of only one spinor $\ket{z}$, starting from the definition \eqref{su2_element_z} of $g(z)$, the normalized Haar measure on $\SU(2)$ is simply the measure induced on the 3-sphere by the Lebesgue measure on $\C^2\sim\R^4$. It takes the form $dg := \frac{1}{2\pi^2}d^2z^0 d^2z^1 \delta(\braaket{z}{z}-1) =: \frac{1}{2\pi^2}d^4z \delta(\braaket{z}{z} - 1 )$, where $2\pi^2$ is the volume of the 3-sphere. Furthermore the group element is invariant under real rescalings of the spinor, $g(\lambda z) = g(z), \, \lambda \in \R$. Therefore, we can write
\beq
\int\limits_{\SU(2)} dg f(g)
& = &
\frac{1}{2\pi^2}\int d^4z \delta(\braaket{z}{z}-1) f(g(z))\nn\\
& = &
2 \int\limits_0^\infty d\lambda \lambda^3 e^{-\lambda^2}
\frac{1}{2\pi^2}\int d^4z \delta(\braaket{z}{z}-1) f(g(\lambda z)) \nn \\
& = &
\frac{1}{\pi^2}\int\limits_0^\infty d\lambda \lambda^3 e^{-\lambda^2}  \int \frac{d^4z}{\lambda^4} \delta\left(\frac{\braaket{z}{z}}{\lambda^2}-1\right)\,f(g(z)) \nn \\
& = &
\frac{1}{\pi^2}  \int\limits d^4z \int d\lambda e^{-\lambda^2}\delta(\lambda - \sqrt{\braaket{z}{z}})\frac{1}{2} f(g(z))    \nn \\
& = &
\frac{1}{2\pi^2} \int d^4z e^{-\braaket{z}{z}} f(g(z)) \,.
\eeq
We simply multiplied by an additional decoupled normalized integral $2 \int_0^\infty d\lambda \lambda^3 \exp(-\lambda^2) = 1$ in the second line and used the invariance of the group element under coordinate-rescaling. In the third line we performed a change of coordinates. The $\delta$-distribution can then be integrated as a $\delta$-distribution in $\lambda$ , which leads to the final expression.

Using the fact that the Haar measure is normalized and invariant under right multiplication, we can come back to $g(z,\tz)$ and recover the result of \cite{spinor2}:
\be
\int_{\SU(2)}dg f(g) = \frac{1}{(2\pi^2)^2}\int\limits_{\C^2 \times \C^2} dz d\tz e^{-\braaket{z}{z} - \braaket{\tz}{\tz}} f(g(z,\tz))\,.
\ee
Note however, that this form of the measure on $\SU(2)$ is by no means unique due to the introduction of redundant degrees of freedom. The integral $2\int d\lambda \lambda^3 \exp(-\lambda^2)$ in the above derivation could indeed be replaced by any other choice of positive and normalized measure. This would lead to a different form of the Haar measure in terms of spinors. The advantage of the above choice is the Gaussian form when written in spinors, which simplifies many computations.

We can follow the same reasoning for the full $\SL(2,\C)$ case. We decompose $G(t,u,\tt,\tu)=G(t,u)\eps^{-1}G(\tt,\tu)^{-1}$, with
\be \label{G_with_u_t}
G(t,u) = \frac{1}{\sqrt{\braaket{u}{t}}}\fourmat{\bar{u}^1}{t^0}{-\bar{u}^0}{t^1}.
\ee
In other words, $G(t,u,\tt,\tu)$ describes the relative $\SL(2,\C)$ transformation between pairs of spinors and, due to the right-invariance of the Haar measure, we can restrict our attention to the group element $G(t,u)$ which depends only on a single pair of spinors. The Haar measure on $\SL(2,\C)$ is the Lebesgue measure on $\C^4\sim\R^8$ with the complex condition $\det G\,=1$,
\beq
\int\limits_{\SL(2,\C)} dG F(G)
& = &
\frac{1}{N}
\int\limits_{\C^4} d^4t d^4u\,
\delta^{(2)}(\braaket{u}{t} - 1)
\,F(G(t,u)) \,.
\eeq
The normalization $N$ should be chosen such that the characters $\chi(G)$ of $\SL(2,\C)$ are orthonormal with respect to this measure. We will not look into this here.
Note that the argument of the $\delta$-distribution is complex in this case.
Now we can use the invariance of $G(t,u)$ under simultaneous complex rescaling:
\be
\ket{t} \rightarrow \lambda \ket{t}, \qquad \ket{u} \rightarrow \bar{\lambda} \ket{u}, \qquad \lambda \in \C \, .
\ee
Introducing another additional, normalized integral $\frac{1}{\cN}\int\limits_\C d^2\lambda |\lambda|^7 \exp(-|\lambda|^2) = 1$ we obtain
\beq
\int \limits_{\SL(2,\C)} dG\,F(G)
& = &
\frac{1}{\cN N} \int\limits_\C d^2\lambda |\lambda|^7 e^{-|\lambda|^2}
\int d^4td^4u\,\delta^{(2)}(\braaket{u}{t} - 1) F(G(\lambda t,\bar{\lambda}u)) \nn \\
& = &
\frac{1}{\cN N} \int\limits_\C d^2\lambda |\lambda|^7 e^{-|\lambda|^2}
\int \frac{d^4t}{|\lambda|^4}\frac{d^4u}{|\lambda|^4} \,
\delta^{(2)}\left(\frac{\braaket{u}{t}}{\lambda^2} - 1\right) F(G(t,u)) \nn \\
& = &
\frac{1}{\cN N}
\int {d^4t}{d^4u} \,
 \int\limits_\C d^2\lambda  e^{-|\lambda|^2}
 \f14(\delta^{(2)}(\lambda-\sqrt{\la u|t\ra})+\delta^{(2)}(\lambda+\sqrt{\la u|t\ra})
F(G(t,u)) \nn \\
& = & \frac{1}{2\cN N} \int d^4u d^4t\,e^{-|\braaket{u}{t}|} F(G(t,u)) \nn  \\
& = & \frac{1}{2\cN N} \int d^4u d^4t\, e^{-\sqrt{A^2 + B^2}} F(G(t,u)) \, ,
\eeq
where we have followed the same steps as in the $\SU(2)$ case above, adapted to complex  rescalings by $\lambda$. Finally we have written the exponential measure factor in terms of the $\SL(2,\C)$ Casimir-invariants described earlier. Once again, we have freedom in changing the normalized integral in $\lambda$ which we inserted, and we chose it here in order to get the simplest expression at the end.

\section{Conclusions and Outlook}

We have shown that the twistor space $\T_2\cong \C^8$, of pairs of twistors $\psi=(\ket{t}, \ket{u})$,
$\tl\psi=(\ket{\tt}, \ket{\tu})$, can be reduced to $T^*\slc$ imposing the area matching constraint
$\bra{u}t\ra\equiv \bra{\tu}\tt\ra$.
The parameterization of Lorentzian holonomy-flux variables in terms of the four spinors is given by \Ref{defJK} and \Ref{Group_element}, which we report here for convenience,
\be
\vec{J} = {\rm Re} \braaaket{t}{\vec{\sigma}}{u}, \qquad \vec{K} = i \, {\rm Im} \braaaket{t}{\vec{\sigma}}{u},
\qquad 
\vec{\tl J} = {\rm Re} \braaaket{\tt}{\vec{\sigma}}{\tu}, \qquad \vec{\tl K} = i \,{\rm Im} \braaaket{\tt}{\vec{\sigma}}{\tu},
\ee
and
\be
G = \Group{t}{\tt}{u}{\tu}, \qquad {\rm such \ that} \qquad 
\Big(\tl J^R, \tl J^L\Big) = \Big(-G^{-1}J^R G, -G^\dagger J^L (G^\dagger)^{-1}\Big).
\ee

The above relation between left- and right-invariant vector fields can be solved for $G$, providing an expression in terms of the normalized bivectors, plus an angle $\Xi$. This is given by \Ref{ganzo}, and generalizes the analogue expression for SU(2), see \Ref{defg}, which is at the roots of the interpretation of semi-classical SU(2) spin networks as a collection of polyhedra.
In particular, $\Xi$ carries information on the 4-dimensional dihedral angle.

Finally, we have presented an action principle encapsulating the whole phase space structure and its constraints, on an arbitrary graph. The action is given by \Ref{action} in terms of the spinors, and by \Ref{action1} in a ``first order formalism'' with both spinors and group elements.
These structures define a notion of twistor networks: an equivalence class of pairs of twistors on each edge, subjected to $U(1)^\C$ area matching conditions on the edges, and $\slc$ gauge invariance conditions on the vertices, and invariant under the generated transformations, that is rescalings on the edges and $\SL(2,\C)$ Lorentz transformations at the vertices.

Within these objects, we identified a class of particular relevance for loop quantum gravity, formed by those twistor networks entirely determined by an SU(2) spinor network and an assignment of boosts on vertices. This class is the classical analogue of the simple projected spin networks living on the boundary of the EPRL/FK models \cite{lift,covariance}.

\section*{Acknowledgements}

The authors are partially supported by the ANR ``Programme Blanc" grants LQG-09.

\appendix
\section{Spinorial Representation of $\sl(2,\C)$} \label{spinorial_appendix}

For the reader's convenience, we summarize in this Appendix our notations and conventions for the Lorentz algebra.
The spinorial representation of $\sl(2,\C)$ can be constructed from the Clifford algebra $C\ell(1,3)$, defined through the anti-commutation-relation of its elements,
$
\{ \gamma^I , \gamma^J \}_A = -\eta^{IJ} 1 , \qquad I,J = 0,\ldots 3.
$
Here $\eta^{IJ}$ is the Minkowski-metric with signature $(-,+,+,+)$, $1$ is the unit-element in $C\ell(1,3)$, the anticommutator is $\{ a, b \}_A := a b + b  a \; \forall a,b \in C\ell(1,3)$. Every irreducible matrix-representation of $C\ell(1,3)$ lives on the space of $4\times 4$-matrices.
Using the commutator $[a,b] : = a b - b a $, one defines
$
\cJ^{IJ} := ({i}/{4}) [\gamma^I, \gamma^J ] ,
$
and it is immediate to check that the $\cJ^{IJ}$ form a representation of $\sl(2,\C)$, i.e.
\beq \label{sl2c_algebra}
[ \cJ^{IJ}, \cJ^{KL} ] = -i \left(  \eta^{JK}\cJ^{IL} + \eta^{IL}\cJ^{JK} - \eta^{IK}\cJ^{JL} - \eta^{JL} \cJ^{IK} \right) \, .
\eeq
One can split these generators of $\sl(2,\C)$ into rotations and boosts according to
\beq
\cJ^i := \frac{1}{2}\epsilon^{ijk}\cJ^{ij}, \quad \cK^i := \cJ^{0i}, \quad i,j,k = 1,2,3,
\eeq
and the algebra reads as
\beq
[\cJ^i , \cJ^j] = i \epsilon^{ijk} \cJ^k, \quad [ \cK^i , \cK^j] = -i \epsilon^{ijk} \cJ^k, \quad [ \cK^i , \cJ^j ] = i \epsilon^{ijk} \cK^k \, .
\eeq
Alternatively one can introduce a chiral splitting of the generators into left and right according to the
isomorphism  $\sl(2,\C) \simeq \su(2) \oplus \su(2)$,
\be
\vec{\cJ^L} := \frac{1}{2}(\vec{\cJ} + i \vec{\cK}), \qquad \vec{\cJ^R} := \frac{1}{2}(\vec{\cJ} - i \vec{\cK}),
\ee
with
\be
[\cJ^L_i  ,\cJ^L_j ] = i\epsilon^{ijk}\cJ^L_k , \qquad [\cJ^R_i  ,\cJ^R_j ] = i\epsilon^{ijk}\cJ^R_k , \qquad [\cJ^L_i  ,\cJ^R_j ] = 0 \, .
\ee

The chiral representation of $C\ell(1,3)$ is given by
\be \label{chiral_basis}
\gamma^0 = \fourmat{0}{-\one}{-\one}{0}, \qquad \gamma^i = \fourmat{0}{\s^i}{-\s^i}{0}, \qquad \gamma^5 = \fourmat{\one}{0}{0}{-\one} \, ,
\ee
where $\one$ is the $2\times 2$ identity matrix and $\s^i$ are the Pauli matrices.
In this basis, 
\be \label{generators_weyl}
\cJ^i  = \frac{1}{2}\fourmat{\s^i}{0}{0}{\s^i}, \qquad \cK^i = \frac{i}{2}\fourmat{\s^i}{0}{0}{-\s^i},
\ee
which shows explicitly that we are dealing with the ${\bf(1/2,0)\oplus(0,1/2)}$ irreducible representation of a Dirac bi-spinor.\\
That $\ket{u}$ and $\ket{t}$ are indeed the left- and right-handed parts of the full Dirac spinor $\psi = \twovec{\ket{t}}{\ket{u}}$ can easily be seen by computing the chiral projectors in this representation:
\be
P^L := \frac{\one_4 - \gamma^5}{2} = \fourmat{0}{0}{0}{\one}, \qquad  P^R := \frac{\one_4 + \gamma^5}{2} = \fourmat{\one}{0}{0}{0} \, .
\ee
Finally, the twistor Poisson brackets \Ref{PBtu} can be compactly written
\be \label{symplectic_weyl}
\{ \psi^\alpha, (\psi^\dagger)^\beta \} = i (\gamma^0)^{\alpha \beta}, \quad \alpha, \beta = 0,1,2,3\, .
\ee

\section{Different spinor representations} \label{Dirac_appendix}
The relation between the two sets of 2-spinors $(\ket{t}, \ket{u})$ and $(\ket{z}, \ket{w})$ can be understood as a 
change of basis for the elements of $C\ell(1,3)$, given by
\be
U := \frac{1}{\sqrt{2}} \fourmat{\one}{\one}{i \one}{-i \one}, \qquad  U^\dagger = U^{-1} =  \frac{1}{\sqrt{2}}\fourmat{\one}{-i\one}{\one}{i\one} \, .
\ee
This gives a new representation (unitarily equivalent) $\hat{\gamma}^I:= U \gamma^I U^\dagger$, with
\beq \label{gamma_matrices}
\hat{\gamma}^0 := \fourmat{-\one}{0}{0}{\one}, \qquad \hat{\gamma}^i := \fourmat{0}{i\s^i}{i\s^i}{0} \qquad \hat{\gamma}^5 := i \hat{\gamma}^0 \hat{\gamma}^1 \hat{\gamma}^2 \hat{\gamma}^3 = i\fourmat{0}{-\one}{\one}{0},
\eeq
and generators 
\beq
\hat{\cJ}^i = \frac{1}{2}\fourmat{\s^i}{0}{0}{\s^i}, \quad \hat{\cK}^i = \frac{1}{2} \fourmat{0}{\s^i}{-\s^i}{0}.
\eeq
This unitary transform acts on the bi-spinors as
$
\hat{\psi} := U \psi,
$
and using the decomposition of a bi-spinor in two spinors $\psi=(\ket{t},\ket{u})$, this becomes
\be
\twovec{\hat{\psi}^0}{\hat{\psi}^1} = \frac{1}{\sqrt{2}}(\ket{t} + \ket{u})\equiv\ket{z}, 
\qquad \twovec{\hat{\psi}^2}{\hat{\psi}^3} = \frac{i}{\sqrt{2}}(\ket{t} - \ket{u})\equiv \ketr{w}.
\ee
Furthermore, the Poisson structure \Ref{symplectic_weyl} changes to
\be \label{symplectic_twistor}
\{ \hat{\psi}^\alpha, (\hat{\psi}^\dagger)^\beta := i (\hat{\gamma^0})^{\alpha \beta} ,
\ee
consistently with \Ref{symplectic_z_w}.


\end{document}